\documentclass[11pt,a4paper]{article}
\usepackage{jcappub} 
\usepackage{rotating} 
\usepackage{graphicx,epsfig}
\usepackage{amsmath}
\usepackage {amssymb}
\usepackage{subfigure}
\pdfoutput=1

\usepackage{relsize}

\providecommand{\U}[1]{\protect\rule{.1in}{.1in}}

\newcommand{\newc}{\newcommand}

\newc{\be}{\begin{equation}}
\newc{\ee}{\end{equation}}

\newc{\ba}{\begin{eqnarray}}
\newc{\ea}{\end{eqnarray}}
\newc{\bea}{\begin{eqnarray*}}
\newc{\eea}{\end{eqnarray*}}
\newc{\D}{\partial}
\newc{\ie}{{\it i.e.} }
\newc{\eg}{{\it e.g.} }
\newc{\etc}{{\it etc.} }
\newc{\etal}{{\it et al.}}
\newc{\lcdm}{$\Lambda$CDM }

\newc{\ra}{\Rightarrow}

\usepackage{cleveref}

\usepackage{slashed}
\usepackage{mathtools}
\usepackage{textcomp}
\usepackage{physics}
\usepackage{soul}
\soulregister\cite7
\soulregister\ref7

\usepackage{parskip}

\usepackage{feynmp-auto}

\usepackage{tcolorbox}

\usepackage{graphicx}  
\usepackage{booktabs}  
\usepackage{array}     
\usepackage{tabularx} 
\usepackage {subfigure} 
\usepackage{tipa}
\usepackage{amsmath,mathrsfs,amsfonts,amssymb}

\renewcommand{\arraystretch}{2.0} 

\usepackage{tikz}
\usepackage{pgfplots}
\pgfplotsset{compat=1.18}
\usetikzlibrary{3d,patterns,arrows.meta}
\usepackage{amssymb}

\title{Decoupling perturbations from background in $f(Q)$ gravity: the square-root 
correction and the impact on the  $\sigma_8$ tension}

\author[a, b]{Chunyu Li,}
\author[ c, d,a, b]{Xin Ren,}
\author[a, b]{Yuhang Yang,}
\author[e,b,f]{Emmanuel N. Saridakis,}
\author[a, b]{Yi-Fu Cai}

\affiliation[a]{Department of Astronomy, School of Physical Sciences, University of Science and Technology of China, Hefei 230026, China}

\affiliation[b]{CAS Key Laboratory for Research in Galaxies and Cosmology, School of Astronomy and Space Science, University of Science and Technology of China, Hefei 230026, China}

\affiliation[c]{Lanzhou Center for Theoretical Physics, Key Laboratory of Theoretical Physics of Gansu Province,
and Key Laboratory of Quantum Theory and Applications of MoE, Lanzhou University, Lanzhou, Gansu 730000, China}
\affiliation[d]{Institute of Theoretical Physics \& Research Center of Gravitation, School of Physical Science and Technology, Lanzhou University, Lanzhou 730000, China}

\affiliation[e]{National Observatory of Athens, Lofos Nymfon 11852, Greece}
\affiliation[f]{Departamento de Matem\'{a}ticas, Universidad Cat\'{o}lica del Norte, Avda. Angamos 0610, Casilla 1280, Antofagasta, Chile}

\emailAdd{springrain@mail.ustc.edu.cn}
\emailAdd{rx76@ustc.edu.cn}
\emailAdd{yyh1024@mail.ustc.edu.cn}
\emailAdd{msaridak@noa.gr}
\emailAdd{yifucai@ustc.edu.cn}

\abstract{We investigate a perturbation-level modification of symmetric teleparallel gravity of the form $f(Q)=F(Q)+M\sqrt{Q}$ and assess its potential to ease the $\sigma_8$ tension. The square-root term leaves the background expansion unchanged at the level of FLRW cosmology, while modifying the effective gravitational coupling and thereby providing a decoupling between background evolution and structure-growth. Using the latest redshift-space distortion data, including DESI DR1 Full-Shape measurements, we constrain the model through its impact on the growth observable $f\sigma_8(z)$ across three representative backgrounds: $\Lambda$CDM, an 
$H_0$-tension-reducing 
model, and a DESI-motivated dynamical dark energy scenario. In all cases, the square-root correction suppresses the growth of structure and induces a degeneracy with $\sigma_8$, leading to weaker constraints from current data. This allows for a wider range of $\sigma_8$ values consistent with Planck, with the effect being most pronounced in the $H_0$-tension-oriented background model. However, because RSD data constrain $f\sigma_8$ rather than $\sigma_8$ directly, a residual 
degeneracy between $M$ 
and $\sigma_8$ remains, indicating that future multi-probe analyses combining 
lensing and full-shape 
clustering will be required to determine whether the $\sqrt{Q}$ term represents 
a genuine signal of 
modified gravity.} 


\begin{document}
\maketitle

\section{Introduction}

The standard cosmological model, namely the $\Lambda$CDM paradigm, provides an 
excellent description of the Universe's evolution across a wide range of 
observations. Nevertheless, several persistent tensions have emerged as data 
precision has improved 
\cite{CosmoVerseNetwork:2025alb,Perivolaropoulos:2021jda}. The well-known $H_0$ 
tension \cite{DiValentino:2020zio} highlights the discrepancy between the Hubble 
constant derived from early-universe Cosmic Microwave Background (CMB) 
measurements and late-universe supernova observations. Analogously, the $S_8$ 
tension \cite{DiValentino:2020vvd} manifests as a discrepancy in the parameter
$    S_8 = \sigma_8 \sqrt{\Omega_m/0.3}$,
where $\sigma_8$ denotes the amplitude of matter fluctuations at the scale of 
$8 h^{-1}\,{\rm Mpc}$ and $\Omega_m$ represents the matter density. Planck CMB 
measurements infer $S_8=0.834 \pm 0.016$ \cite{Planck:2018vyg}, whereas 
large-scale structure (LSS) observations from galaxy clustering and weak 
lensing surveys consistently prefer lower values \cite{Joudaki:2016kym, 
Hildebrandt:2016iqg, DES:2017qwj, Heymans:2020gsg, KiDS:2020suj, 
Troster:2019ean}, resulting in a $2$-$3\sigma$ tension. 
Ref.~\cite{DiValentino:2020vvd} demonstrated that this discrepancy is primarily 
driven by $\sigma_8$ rather than $\Omega_m$. Consequently, in this work we 
focus on the $\sigma_8$ (or, equivalently, $S_8$) tension.

Modified gravity  theories \cite{CANTATA:2021asi} provide a natural framework to 
address such cosmological tensions by altering both the background expansion 
history and the evolution of perturbations. Within this context, symmetric 
teleparallel gravity based on the non-metricity scalar $Q$, known as $f(Q)$ 
gravity \cite{BeltranJimenez:2017tkd, BeltranJimenez:2019tme}, has emerged as a 
theoretically well-motivated candidate, with interesting cosmological 
phenomenology  \cite{Anagnostopoulos:2021ydo,Lazkoz:2019sjl,Lu:2019hra,
  Mandal:2020buf,  
  Gadbail:2022jco, 
Khyllep:2021pcu,Hohmann:2021ast,Mandal:2021bpd,Capozziello:2022wgl,  
Solanki:2022ccf, Narawade:2022jeg,Capozziello:2022tvv, Shabani:2023xfn, capozziello2022comparing, 
Ferreira:2022jcd, Arora:2022mlo,
Paliathanasis:2023nkb,DAgostino:2022tdk,De:2022jvo,BeltranJimenez:2022azb,
De:2023xua,Koussour:2023rly,  Boehmer:2023knj,Bhagat:2023ych,Gadbail:2023mvu,
Shabani:2023nvm,Lohakare:2023ugg,Paliathanasis:2023pqp, Shi:2023kvu,Hu:2023ndc,
Dimakis:2021gby,Anagnostopoulos:2022gej,Lymperis:2022oyo,Yang:2024tkw, 
DAmbrosio:2023asf,Guzman:2024cwa,Capozziello:2024jir,  Basilakos:2025olm} (for a 
 review see 
\cite{Heisenberg:2023lru}). 
Symmetric teleparallel gravity includes General Relativity  as a specific limit 
and has the advantage of  second-order field equations. Previous studies have 
demonstrated that $f(Q)$ models can effectively alleviate cosmological tensions 
\cite{Sakr:2024eee, Wang:2024eai, Boiza:2025xpn, Najera:2025htf, 
Kavya:2025vsj}. However, late-time extensions that increase $H_0$ typically 
reduce $\sigma_8$ simultaneously, thereby exacerbating the growth tension, as 
discussed in \cite{Heisenberg:2022gqk}. This feature suggests that a viable 
framework should be able to modify $\sigma_8$ and $H_0$ in a controlled and, 
ideally, partially independent manner.

Working within  the general framework of $f(Q)$ gravity, we focus on the 
modification class
\begin{equation}
    f(Q) = F(Q) + M\sqrt{Q}\,, \nonumber
\end{equation}
as previously considered in \cite{Barros:2020bgg, Frusciante:2021sio, 
Atayde:2021pgb, Kolhatkar:2025ubm}, where the parameter $M$ has units of 
$[L^{-1}]$. The inclusion of the $\sqrt{Q}$ correction introduces a new degree 
of freedom with a distinctive and, for our purposes, crucial property: for 
homogeneous and isotropic backgrounds, this term cancels out of the modified 
Friedmann equations, leaving the background expansion history unchanged and 
thus preserving constraints from background observables. Beyond FLRW background (such as perturbative or anisotropic configurations), this cancellation no longer holds. At the level of linear perturbations, the $\sqrt{Q}$ term modifies the effective gravitational coupling $G_{\mathrm{eff}}$, thereby affecting the growth of structures. As for higher orders, it changes the non-linear kernels, impacting higher-order statistics such as the bispectrum and the three-point correlation function. Therefore, this   
decoupling between background and linear perturbation dynamics is highly non-generic 
in modified gravity theories and makes the model particularly suitable as a 
successful theoretical laboratory to investigate late-time clustering tensions in large-scale structure data. This property allows for a more independent variation of $\sigma_8$ and $H_0$: while $H_0$ is mainly controlled by the background expansion through the $F(Q)$, $\sigma_8$ is affected through the $G_{\mathrm{eff}}$. This may accommodate higher $H_0$ values without the corresponding suppression of $\sigma_8$ usually encountered in related modified-gravity scenarios \cite{Heisenberg:2022gqk}.

Given that the $M\sqrt{Q}$ correction specifically alters the evolution of 
perturbations without affecting the background dynamics, it is crucial to 
employ observational probes that are directly sensitive to structure formation. 
Redshift-space distortions (RSD) \cite{Kaiser:1987qv} are ideally suited for 
this purpose, as they constrain the growth rate of structure through the 
combination $f\sigma_8(z)$. Since RSD measurements provide a direct 
observational handle on density perturbations, they are expected to yield tight 
constraints on the $\sqrt{Q}$ term and thus offer a robust test of the model's 
capability to alleviate the $\sigma_8$ tension \cite{Barros:2018efl, 
Lambiase:2018ows, Gomez-Valent:2018nib, Yan:2019gbw, SolaPeracaula:2020vpg, 
Boiza:2025xpn}. At the same time, because RSD measurements constrain $f\sigma_8$ rather than $\sigma_8$ directly, an important degeneracy between $M$ and $\sigma_8$ is expected when RSD data are used alone. In particular, the recent DESI DR1 Full-Shape analysis \cite{DESI:2024jxi} provides the most precise low-redshift measurements of $f\sigma_8$ currently available, modestly reducing statistical uncertainties compared to previous compilations.

Previous studies have explored related scenarios in $f(Q)$ gravity. 
Ref.~\cite{Barros:2020bgg} constrained the parameters of the $Q+M\sqrt{Q}$ 
model using RSD data, while Refs.~\cite{Frusciante:2021sio, Atayde:2021pgb, 
Sakr:2024eee} combined multiple cosmological probes to obtain broader 
constraints on $f(Q)$ models. Crucially, these earlier analyses of the $\sqrt{Q}$ correction tested the $M\sqrt{Q}$ term within a single background cosmology. In the present work we advance these efforts in 
several directions. Firstly, we exploit the latest $f\sigma_8$ measurements 
from DESI Full-Shape clustering \cite{DESI:2024jxi} together with a robust 
compilation of RSD data. This leads to an improvement in statistical uncertainties relative to earlier RSD datasets and allows for a consistency check of the inferred constraints, as we jointly constrain the $M\sqrt{Q}$ term and 
$\sigma_8$, thereby directly testing the hypothesis that a perturbation-only 
modification can reconcile LSS data with Planck-inferred clustering amplitudes. 
Secondly, to test the robustness of this mechanism beyond a single-background setup, we investigate the interplay between the $M\sqrt{Q}$ correction and 
different expansion histories by embedding the same $\sqrt{Q}$ perturbation 
sector into three distinct $f(Q)$ background models: (i) a model that 
reproduces the standard $\Lambda$CDM background, (ii) an exponential model that 
has been shown to alleviate the $H_0$ tension, and (iii) a quadratic model 
realizing quintom dynamical dark energy, motivated by recent DESI results indicating 
that dark energy may evolve over time \cite{DESI:2024mwx, DESI:2025zgx}. This 
strategy allows us to assess to what extent a single perturbative degree of 
freedom can mitigate the $\sigma_8$ tension across qualitatively different 
background cosmologies. Neglecting the time derivatives of the gravitational potentials and assuming $aH \ll k$, i.e. the 
regime relevant to our analysis, the predictions of $f(Q)$ and $f(T)$ models 
coincide \cite{BeltranJimenez:2017tkd, BeltranJimenez:2019tme}, thus our 
results also have implications for torsional modified gravity.

The remainder of the paper is organized as follows. In Sec.~\ref{sec:the}, we 
review the theoretical framework of $f(Q)$ gravity and outline the perturbation 
theory relevant for structure growth. In Sec.~\ref{sec:data}, we describe the 
observational datasets and the fitting methodology used in our analysis. 
Sec.~\ref{sec:result} presents our parameter constraints and discusses the 
implications for the $\sigma_8$ tension and the role of the $M\sqrt{Q}$ term. 
Finally, Sec.~\ref{sec:conc} summarizes our conclusions and outlines possible 
directions for future tests.

\section{ $f(Q)$ gravity and cosmology}
\label{sec:the}

In this work we investigate modified gravity within the framework of $f(Q)$ 
theories, 
where gravitational dynamics are governed by the non-metricity scalar $Q$ 
rather than curvature or torsion 
\cite{BeltranJimenez:2017tkd, BeltranJimenez:2019tme}. 
This section provides a concise overview of the geometrical foundations, 
background evolution, 
and perturbation dynamics relevant to our analysis, with particular emphasis on 
the distinctive role 
played by the square-root correction $M\sqrt{Q}$.

\subsection{Geometry and field equations in $f(Q)$ 
gravity}

The fundamental quantity of symmetric teleparallel gravity is the non-metricity 
tensor 
\begin{equation}
    Q_{\alpha\mu\nu} = \nabla_\alpha g_{\mu\nu}\ ,
\end{equation}
which implies that the vector lengths are not preserved under parallel 
transportations. 
From this tensor one constructs the non-metricity scalar
\begin{equation}
    Q = - Q_{\alpha\mu\nu} P^{\alpha\mu\nu}\ ,
\end{equation}
where the superpotential (or non-metricity conjugate)
\begin{equation} \label{conjugate}
P^{\alpha}\,_{\mu\nu} = -\frac{1}{2}L^{\alpha}{}_{\mu\nu}
+\frac{1}{4}(Q^{\alpha}-\tilde{Q}^{\alpha})g_{\mu\nu}
-\frac{1}{4}\delta^{\alpha}_{(\mu}Q_{\nu)}\ ,
\end{equation}
is constructed from
\begin{equation}
    L^{\alpha}{}_{\mu\nu} = \frac{1}{2} Q^{\alpha}{}_{\mu\nu} - 
Q_{(\mu\nu)}{}^{\alpha}\,,
\end{equation}
with the two traces
\begin{equation}
Q_\alpha = g^{\mu\nu} Q_{\alpha\mu\nu}\,, 
\qquad 
\tilde{Q}_\alpha = g^{\mu\nu} Q_{\mu\alpha\nu}\,.
\end{equation}

Hence, the   action of $f(Q)$ gravity reads
\begin{equation}
\label{eq:action}
\mathcal{S} = \int d^4 x \sqrt{-g}
\left[ -\frac{1}{16\pi G} f(Q) + \mathcal{L}_m \right],
\end{equation}
where $\mathcal{L}_m$ is the matter Lagrangian density.

\subsection{Background dynamics}
\label{sec:background}

We adopt a spatially flat Friedmann-Lemaître-Robertson-Walker (FLRW) metric 
\cite{BeltranJimenez:2017tkd} 
\begin{equation}
ds^2 = -dt^2 + a^2(t)\delta_{ij}dx^i dx^j,
\end{equation}
for which the non-metricity scalar reduces to
\begin{equation}
Q = 6H^2\,, \qquad H=\dot{a}/a.
\end{equation}

Variation of the action~\eqref{eq:action} yields the modified Friedmann 
equations 
\begin{equation}
\label{eq:fldm1}
6 f_Q H^2 - \frac{f}{2} = 8\pi G \rho_m\ ,
\end{equation}
\begin{equation}
\label{eq:fldm2}
(12H^2 f_{QQ} + f_Q)\dot{H} = -4\pi G(\rho_m + p_m)\,,
\end{equation}
with $\rho_m$ and $p_m$ the energy density and pressure of the matter sector, 
assumed to correspond to an ideal fluid.

Let us now consider the form
\begin{equation}
\label{formweuse}
f(Q) = F(Q) + M\sqrt{Q}\,.
\end{equation}
As one can see by inserting (\ref{formweuse}) into the Friedmann equations, 
the square-root term cancels identically in both of them. We show this explicitly in Appendix \ref{appA}.
Thus,  the background evolution, namely the cosmological expansion history,
depends solely on $F(Q)$, while $M$ affects only perturbations.  
This feature, which only holds on isotropic and homogeneous backgrounds, is non-generic among modified gravity theories and is central to 
our study. Finally, we mention that   imposing $F(Q)=Q+2\Lambda$ yields a 
$\Lambda$CDM background.

\subsection{Linear perturbations and the effective gravitational coupling}

Although the term $M\sqrt{Q}$ leaves the background invariant, it modifies the 
evolution of scalar perturbations.  
In particular, in the Newtonian gauge 
\begin{equation}
    ds^2 = -(1+2\Psi)dt^2 + a^2(1-2\Phi)\delta_{ij}dx^i dx^j,
\end{equation}
and, in the regime where the time derivatives of the gravitational potentials are neglected and $aH \ll k$, one has $\Phi=\Psi$ and the evolution is scale-independent. Outside this regime, scale dependence is generally expected to appear. In this case, the modified Poisson equation becomes \cite{BeltranJimenez:2019tme}
\begin{equation}
-k^2\Psi = 4\pi \frac{G_N}{f_Q} a^2 \rho_m \delta\,,
\end{equation}
with $\delta\equiv\delta\rho_m/\rho_m$ the matter overdensity.

The matter perturbation growth equation is
\begin{equation}
\label{eq:fqper}
\ddot{\delta} + 2H\dot{\delta}
- \frac{4\pi G_N}{f_Q}\rho_m \delta = 0\,,
\end{equation}
implying an effective gravitational coupling of the form
\begin{equation}
G_{\rm eff} = \frac{G_N}{f_Q}\ ,
\end{equation}
analogous to $f(T)$ models \cite{Zheng:2010am, Bahamonde:2020lsm}. 
Since
\begin{equation}
f_Q = F_Q + \frac{M}{2\sqrt{Q}}\,,
\end{equation}
a positive $M$ increases $f_Q$ and therefore reduces $G_{\rm eff}$, producing a 
weaker effective gravity on growth scales, and thus suppressing structure 
formation, while 
a negative $M$ enhances it. This modification directly affects matter perturbation $\delta$ and consequently observables such as the growth rate $f$, which can be probed by large-scale structure measurements, e.g., redshift-space distortions. Hence, the parameter $M$ serves as a pure perturbative handle for modifying the growth rate without changing $H(z)$.

\subsection{Specific background realizations}
\label{sec:model}

In this subsection we summarize the three illustrative background choices 
used in our analysis. All models lie within the well-studied classes of 
symmetric teleparallel theories with no pathologies or instabilities.  
Each is compatible with the same perturbation-sector modification through 
$M\sqrt{Q}$.

\subsubsection{Model A: $\Lambda$CDM background}

The simplest choice,
\begin{equation}
F_A(Q) = Q + 2\Lambda\,,
\end{equation}
yields $f(Q)=Q+M\sqrt{Q}+2\Lambda$.  Additionally, note that 
equation \eqref{eq:fqper} reduces to the general-relativity form when $M=0$. 
We can re-write (\ref{eq:fqper}) using $N=\ln a$ as the independent variable, 
namely 
\begin{equation}
\delta^{\prime\prime}+\delta^{\prime}\left(2+\frac{H^{\prime}}{H}\right)-\frac{
3\sqrt{6}H}{2\sqrt{6}H+M}\Omega_m\delta=0\ ,
\end{equation}
where primes denote derivatives with respect to $N$ (hence 
$\dot{\delta}=H\delta'$) and 
where we have introduced the matter density parameter  $\Omega_m = 8\pi 
G\rho_m/(3H^2)$.

\subsubsection{Model B: $H_0$-tension alleviating background}

A frequently studied form capable of enhancing late-time expansion 
\cite{Anagnostopoulos:2021ydo,Boiza:2025xpn} is
\begin{equation}
F_B(Q) = Q e^{\beta Q_0/Q},
\end{equation}
with $\beta$ and $Q_0$ the model parameters.
The normalized Hubble function $E^2 = H^2/H_0^2$ (from now one the subscript 
``0'' denotes the value of a quantity at present) satisfies
\begin{equation}
(E^2 - 2\beta)e^{\beta/E^2} = \Omega_{m0} a^{-3}\,,
\end{equation}
with
\begin{equation}
\beta = \frac{1}{2} + W_0\left(-\frac{\Omega_{m0}}{2\sqrt{e}}\right),
\end{equation}
where $W_0$ denotes the principal branch of the Lambert function.  
This background evolution approaches general relativity at high redshift (i.e. 
at $Q\gg Q_0$), but it does not recover  $\Lambda$CDM scenario at late times,  
hence it is a successful alternative to the latter, capable of 
  alleviating the $H_0$ tension \cite{Boiza:2025xpn,Najera:2025htf}.

\subsubsection{Model C: Quintom dark energy background}

Motivated by DESI indications of a quintom-like evolution of dark energy 
\cite{Yang:2024kdo,Yang:2025kgc,roy2025dynamical,gu2025dynamical,goh2025phantom,li2025robust,Yang:2025mws,Lu:2025gki,Lu:2021wif}, we consider the quadratic model
\begin{equation}
\label{eq:quadratic}
F_C(Q) = Q + \alpha_1 Q + \alpha_2 \frac{Q^2}{Q_0} - 2\alpha_3 Q_0\,,
\end{equation}
shown to allow the effective equation-of-state parameter $w(z)$ to cross $-1$ 
\cite{Feng:2004ad, Huterer:2004ch, Cai:2009zp, Cai:2025mas}.  
This model provides a flexible dynamical dark energy sector, while remaining 
close to $\Lambda$CDM over the redshift range ($z<2$) relevant for our 
growth-rate analysis.

\subsection{Perturbative Stability}
\label{sec:stability}

To assess the physical viability of the models under consideration, we examine three aspects of perturbative stability: ghost stability, gradient stability, and strong coupling.

\subsubsection*{Ghost Stability}

Ghost stability requires that the kinetic terms in the quadratic action are positive definite. In $f(Q)$ gravity on a flat FLRW background, this translates into the conditions $f_Q > 0$ for tensor perturbations and $f_Q + 2Qf_{QQ} > 0$ for scalar perturbations, while vector modes do not introduce propagating degrees of freedom\cite{BeltranJimenez:2019tme}. These conditions are satisfied for all three models within the observationally allowed parameter space.

\subsubsection*{Gradient Stability}

Gradient stability requires a non-negative effective sound speed squared, $c_s^2 \geq 0$, preventing the exponential growth of perturbations on small scales. For tensor and vector perturbations, it has been shown that the conditions for gradient stability coincide with those for ghost stability.

In the scalar sector, although a complete second-order action is not yet available within $f(Q)$ gravity framework, several robust conclusions can be drawn. Since the field equations remain second order, no additional propagating scalar degree of freedom arises beyond those in General Relativity. In the GR limit, the scalar sound speed satisfies $c_s^2 = 1$. For the models considered here, which represent small deviations from $\Lambda$CDM, deviations in $c_s^2$ are expected to remain small, suggesting that the condition $c_s^2 > 0$ is satisfied within the viable parameter space.

\subsubsection*{Strong Coupling and EFT Cutoff}

When the kinetic term of a perturbation mode vanishes in the quadratic action, the corresponding degree of freedom may not appear at the linear level even though it is present in the full theory. In such situations perturbation modes can couple non-trivially to one another, which is generally regarded as a signal of a potential strong coupling problem and signals a breakdown of the perturbative expansion below a certain energy cutoff. For $f(Q)$ gravity on a general FLRW background, the kinetic terms remain non-zero and the linear perturbation analysis is well defined for the models considered here~\cite{BeltranJimenez:2019tme, Hu:2023juh, Zhao:2024kri}. Since our analysis is conducted on a cosmological FLRW background away from these degenerate limits, the strong coupling problem does not affect the perturbative regime relevant to the present work.

\begin{table}[htp]
    \centering
    \renewcommand{\arraystretch}{1.5}
    \setlength{\tabcolsep}{12pt}
    \begin{tabular}{ccccc}
        \hline
        \hline
        $z$     & $f\sigma_8(z)$ & $\sigma_{f\sigma_8}(z)$  & 
$\Omega_{m,0}^\text{ref}$ & Ref. \\ 
        \hline
        0.02    & 0.428 & 0.0465  & 0.3 &  \cite{Huterer:2016uyq}   \\
        0.02    & 0.398 & 0.065   & 0.3 &  \cite{Turnbull:2011ty}, 
\cite{Hudson:2012gt} \\
        0.02    & 0.314 & 0.048   & 0.266 &  \cite{Davis:2010sw}, 
\cite{Hudson:2012gt}  \\
        0.10    & 0.370 & 0.130   & 0.3 &  \cite{Feix:2015dla}  \\
        0.15    & 0.490 & 0.145   & 0.31 &  \cite{Howlett:2014opa}  \\
        0.17    & 0.510 & 0.060   & 0.3 &  \cite{Song:2008qt}  \\
        0.18    & 0.360 & 0.090   & 0.27 &  \cite{Blake:2013nif} \\
        0.38    & 0.440 & 0.060   & 0.27 &  \cite{Blake:2013nif} \\
        0.25    & 0.3512 & 0.0583 & 0.25 &  \cite{samushia2012interpreting} \\
        0.37    & 0.4602 & 0.0378 & 0.25 &  \cite{samushia2012interpreting} \\
        0.32    & 0.384 & 0.095  & 0.274 &  \cite{BOSS:2013eso}   \\
        0.59    & 0.488  & 0.060 & 0.307115 &  \cite{BOSS:2013mwe} \\
        0.44    & 0.413  & 0.080 & 0.27 &  \cite{blake2012wigglez} \\
        0.60    & 0.390  & 0.063 & 0.27 &  \cite{blake2012wigglez} \\
        0.73    & 0.437  & 0.072 & 0.27 &  \cite{blake2012wigglez} \\
        0.60    & 0.550  & 0.120 & 0.3 &  \cite{Pezzotta:2016gbo} \\
        0.86    & 0.400  & 0.110 & 0.3 &  \cite{Pezzotta:2016gbo} \\
        1.40    & 0.482  & 0.116 & 0.27 &  \cite{Okumura:2015lvp} \\
        0.978   & 0.379  & 0.176 & 0.31 &  \cite{eBOSS:2018yfg} \\
        1.23    & 0.385  & 0.099 & 0.31 &  \cite{eBOSS:2018yfg} \\
        1.526   & 0.342  & 0.070 & 0.31 &  \cite{eBOSS:2018yfg} \\
        1.944   & 0.364  & 0.106 & 0.31 &  \cite{eBOSS:2018yfg} \\
        0.295   & 0.37784 & 0.09446 & 0.315 &  \cite{DESI:2024jxi} \\
        0.510   & 0.515879 & 0.061529 & 0.315 &  \cite{DESI:2024jxi} \\
        0.706   & 0.483315 & 0.055236 & 0.315 &  \cite{DESI:2024jxi} \\
        0.930   & 0.422208 & 0.046179 & 0.315 &  \cite{DESI:2024jxi} \\
        1.137   & 0.37468 & 0.037468 & 0.315 &  \cite{DESI:2024jxi} \\
        1.491   & 0.4350  & 0.045 & 0.315 &  \cite{DESI:2024jxi} \\
        \hline
        \hline
    \end{tabular}
    \caption{Compilation of the $f\sigma_8(z)$ measurements used in this 
work.}
    \label{rsddata}
\end{table}

\section{Observational data and methodology}
\label{sec:data}

Redshift-space distortions (RSD) constitute one of the most direct 
observational probes of the growth of cosmic structure. The peculiar velocities 
of galaxies introduce anisotropies in the observed clustering pattern, 
generating apparent distortions along the line of sight. Since these distortions 
are driven by the rate at which density perturbations grow under gravity, RSD 
measurements provide a robust test for probing deviations from general 
relativity and for constraining modified gravity scenarios such as the $f(Q)$ 
models considered here.

\subsection{The observable  $f\sigma_8$} 

As it is known, RSD measurements do not independently constrain the linear 
growth rate $f$ or the clustering amplitude $\sigma_8$, but instead provide 
constraints on the combined observable
\begin{equation}
    f\sigma_8(a)
    = \frac{d\ln\delta(a)}{d\ln a}\,\sigma_8(a)
    = \frac{\sigma_8(1)}{\delta(1)}\,a\, \frac{d\,\delta(a)}{da},
\label{eq:fs8a}
\end{equation}
where $\sigma_8 \equiv \sigma_8(a=1)$ is the 
present-day matter fluctuation amplitude at $8h^{-1}{\rm Mpc}$. 
Expressed in terms of the number of e-folds $N = \ln a$, 
expression \eqref{eq:fs8a} becomes
\begin{equation}
    f\sigma_8(N) = \frac{\delta'(N)}{\delta(N)}\,\sigma_8,
\end{equation}
implying  that any modification to the perturbation equation, such as that 
induced by the $M\sqrt{Q}$ term in $f(Q)$ gravity, directly affects the 
predicted $f\sigma_8$ signal. Given that RSD alone is only sensitive to this combination, a degeneracy between $M$ and $\sigma_8$ is expected in our constraint results, highlighting a key limitation of relying solely on this single probe.

\subsection{RSD Dataset}

The compilation of $f\sigma_8(z)$ measurements used in this work is listed in 
Table~1.  
The first 22 data points correspond to the well-tested, internally robust 
dataset analyzed in~\cite{Sagredo:2018ahx}, while the final six measurements 
are the recent DESI DR1 Full-Shape clustering results~\cite{DESI:2024jxi}. 
Together, they provide extensive redshift coverage and represent the most 
precise growth-rate measurements currently available, enabling stringent tests 
of the $f(Q)$ models under consideration.

\subsection{Alcock-Paczyński Corrections}

The Alcock-Paczyński effect arises when the conversion from observed 
redshifts to comoving distances assumes a fiducial cosmology different from the 
true one. Such geometric distortions induce anisotropies mimicking or modifying 
the RSD signal. If a measurement $f\sigma_8'(z)$ is obtained under a fiducial 
Hubble parameter $H'(z)$, one must apply the correction~\cite{Macaulay:2013swa, 
Kazantzidis:2018rnb,Sagredo:2018ahx}
\begin{equation}
    f\sigma_8(z) \simeq
    \frac{H(z)D_A(z)}{H'(z)D_A'(z)} \, f\sigma_8'(z)
    \equiv q(z,\Omega_{m0},\Omega_{m0}')\,f\sigma_8'(z),
\end{equation}
where the angular diameter distance is
\begin{equation}
    D_A(z) = \frac{1}{1+z}\int_0^z \frac{dz'}{H(z')}.
\end{equation}
For realistic background cosmologies, the correction is mild, typically $\sim 
2\%-3\%$ around $z\sim 1$~\cite{Kazantzidis:2018rnb}.
Hence, to incorporate this correction consistently, we define the residual 
vector
\begin{equation}
    V^i \equiv f\sigma_{8,i}
    - \frac{f\sigma_8(z_i,\Omega_{m0},\sigma_8)}
    {\,q(z_i,\Omega_{m0},\Omega_{m0}^{\text{fid}_i})\,},
\end{equation}
where $f\sigma_{8,i}$ denotes the observed value and $q$ is the  
Alcock-Paczyński factor.

\subsection{Likelihood analysis}

The total chi-squared used for parameter estimation is
\begin{equation}
    \chi^2 = V^i\, C^{-1}_{ij}\, V^j,
\end{equation}
where $C^{-1}_{ij}$ is the inverse covariance matrix. All measurements are 
treated as uncorrelated except for the WiggleZ subsample, for which we use the 
published covariance matrix
\begin{equation}
C^{\rm WiggleZ}_{ij} = 10^{-3}
\begin{pmatrix}
6.400 & 2.570 & 0.000 \\
2.570 & 3.969 & 2.540 \\
0.000 & 2.540 & 5.184
\end{pmatrix}.
\end{equation}

Now, in order to compare the performance of different models, we employ the corrected
Akaike (AIC$_c$) for small data sample sizes~\cite{akaike2003new, Sugiura01011978} and Bayesian 
(BIC)~\cite{schwarz1978estimating} information criteria, characterized by 
\begin{align}
    \mathrm{AIC}_c &= -2\ln L_{\max} + 2N_p+ \frac{2N_p(N_p+1)}{N_d - N_p - 1}, \\
    \mathrm{BIC} &= -2\ln L_{\max} + N_p \ln N_d,
\end{align}
where $L_{\max}$ is the maximum likelihood, $N_p$ is the number of model 
parameters, and $N_d$ is the number of data points.  Then, by calculating the 
differences $\Delta\mathrm{IC}$ relative to the model of reference, we have  
the following possibilities:  $\Delta\mathrm{IC}<2$ implies no 
  preference and the two models are statistically equivalent; $
4\lesssim \Delta\mathrm{IC}\lesssim 7$ implies moderate less support; $
\Delta\mathrm{IC}>10$ implies essentially no support.

To provide a more robust and rigorous model comparison, we also evaluate the Bayesian evidence $\mathcal{Z}$ for each considered model. The evidence is computed using the Bridge sampling method applied directly to the generated MCMC chains, following the approach described in \cite{gronau2017tutorial}. The statistical preference between a given model and the reference model is quantified by the difference in the log-evidence,
\begin{equation}
\Delta \ln \mathcal{Z} = \ln \mathcal{Z}_{\mathrm{model}} - \ln \mathcal{Z}_{\mathrm{ref}} \, .
\end{equation}

According to the revised Jeffreys' scale, the evidence in favor of the model with the higher $\mathcal{Z}$ is interpreted as follows: $0 \leq \Delta \ln \mathcal{Z} < 1$ indicates weak or inconclusive evidence; $1 \leq \Delta \ln \mathcal{Z} < 2.5$ implies moderate evidence; $2.5 \leq \Delta \ln \mathcal{Z} < 5$ implies strong evidence; and $\Delta \ln \mathcal{Z} \geq 5$ constitutes decisive evidence.

\subsection{Models and priors}

As we mentioned above, we analyze three $f(Q)$ models incorporating the same 
perturbation-sector correction $M\sqrt{Q}$, namely 
\begin{eqnarray}
    &&\text{Model A}:\quad f(Q) = Q + 2\Lambda + M\sqrt{Q}, \nonumber \\
    &&\text{Model B}:\quad f(Q) = Q e^{\beta Q_0/Q} + M\sqrt{Q}, \nonumber \\
    &&\text{Model C}:\quad f(Q) = Q + \alpha_1 Q + \alpha_2\frac{Q^2}{Q_0} - 
2\alpha_3 Q_0 + M\sqrt{Q}.
\label{eq:model}
 \end{eqnarray}
For each background, we consider the two cases, namely 
\begin{enumerate}
    \item $M = 0$ (no perturbative modification), 
    \item $M$ free and jointly constrained with $\sigma_8$.
\end{enumerate}
Moreover, we adopt broad, flat priors:
\[
    \sigma_8 \in [0.6,1.1], \qquad 
    M \in [-3H_0,\,11H_0].
\]
The lower bound on $M$ is chosen such that both ghost and gradient stability are ensured. We adopt a broad uniform prior on $M$ to avoid artificially truncating the posterior. An explicit sensitivity test reducing the prior range (e.g., from $M \in [-210, 770]$ to $[-100, 500]$) confirms that our main results and the $M$-$\sigma_8$ degeneracy structure are robust and not driven by the prior width. Note that since   $M$ has units of $[L^{-1}]$, i.e. of $H$, for convenience we measure it in $(km\cdot s^{-1}\cdot Mpc^{-1})$.

We fix the background parameters $(H_0, \Omega_{m0})$ rather than marginalizing over them because RSD data alone yield severe degeneracies between the modified gravity parameter $M$ and both $\Omega_{m0}$ and $H_0$ in the growth equation. Jointly constraining them from RSD data is therefore ill-conditioned. Consequently, to ensure internal consistency, each background uses fiducial values 
$(H_0,\Omega_{m0})$ corresponding to current best-fit constraints:

i) {Model A}: $(68.52,\ 0.2948)$ (DESI 
$\Lambda$CDM)~\cite{DESI:2024mwx},

ii) {Model B}: $(70.431,\ 0.3386)$, consistent 
with~\cite{Boiza:2025xpn},

iii) {Model C}: $(68.03,\ 0.3085)$ (DESI $w_0w_a$CDM).
 
Finally, note that for Model C we further use the  background parameters
$
\alpha_1 = -0.062,  \alpha_2 = 0.003,  \alpha_3 = -0.38$, 
as obtained in~\cite{Yang:2025mws}.

\section{Results and discussion}
\label{sec:result}

\begin{figure*}[t]
\centering
\includegraphics[width=0.5\textwidth]{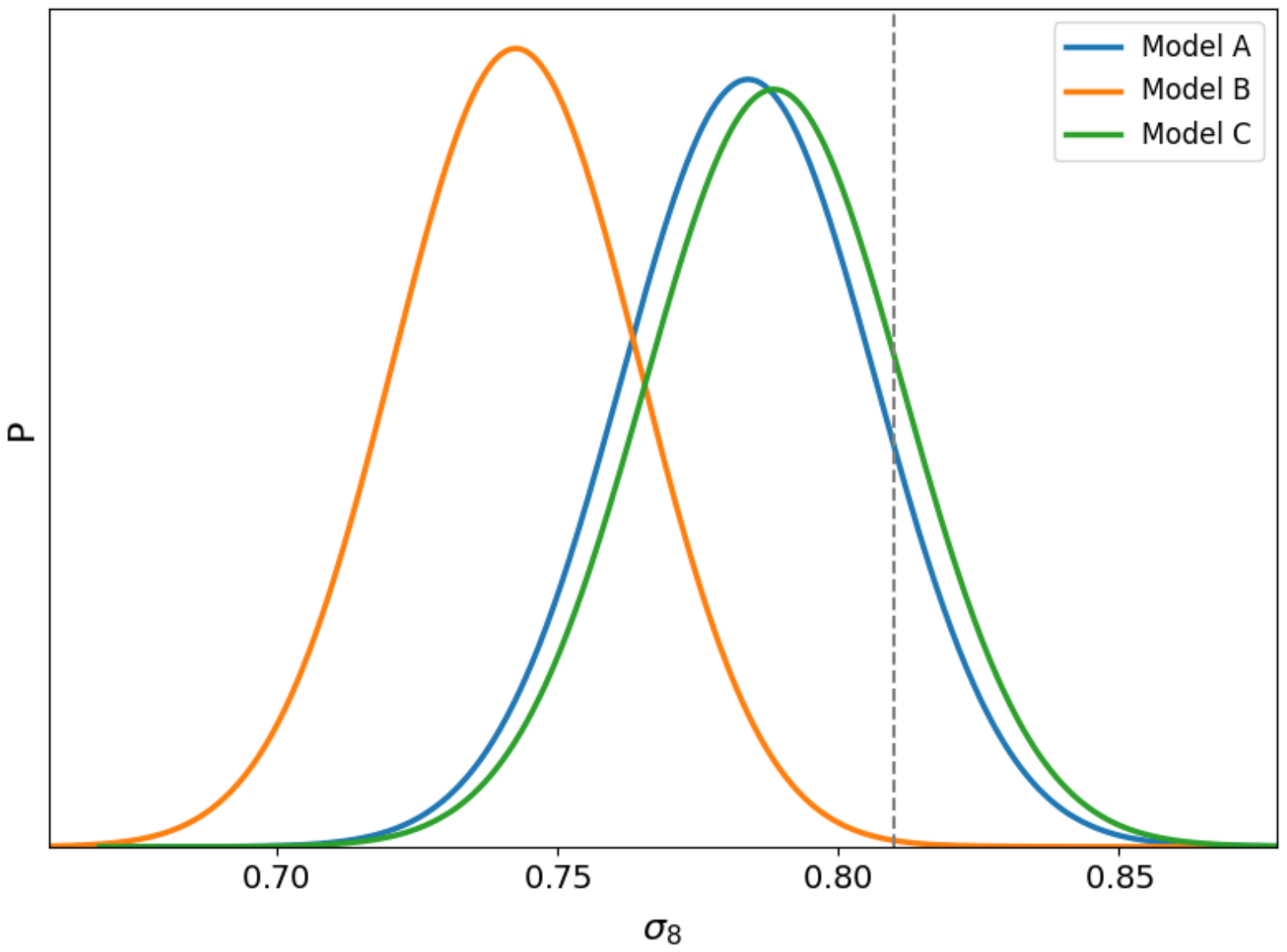}
\includegraphics[width=0.9\textwidth]{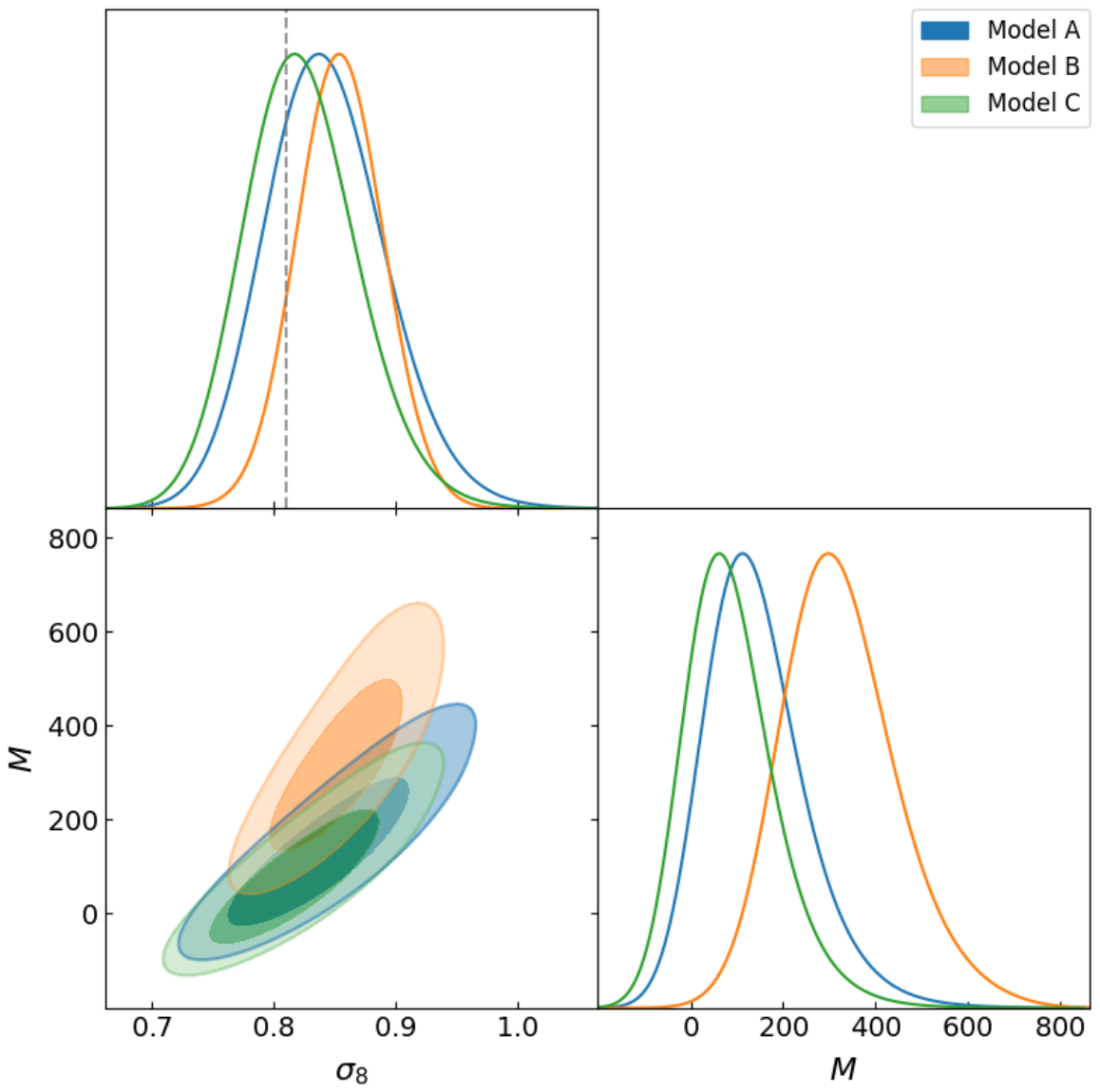}
\caption{ {\it {\textbf{Upper panel}: Constrains for $\sigma_8$ values when $M=0$; \textbf{Lower panel}: Marginalised constraints for $\sigma_8$ and $M$ values at 68\% and 95\% C.L. Models A, B, and C are defined according to equation \eqref{eq:model}. The gray dashed line represents the value $\sigma_8=0.811$, which aligns with the Planck 2018 results.}}}
\label{fig:corner}
\end{figure*}

\begin{figure*}[t]
    \centering
    \includegraphics[width= 1.0 \linewidth]{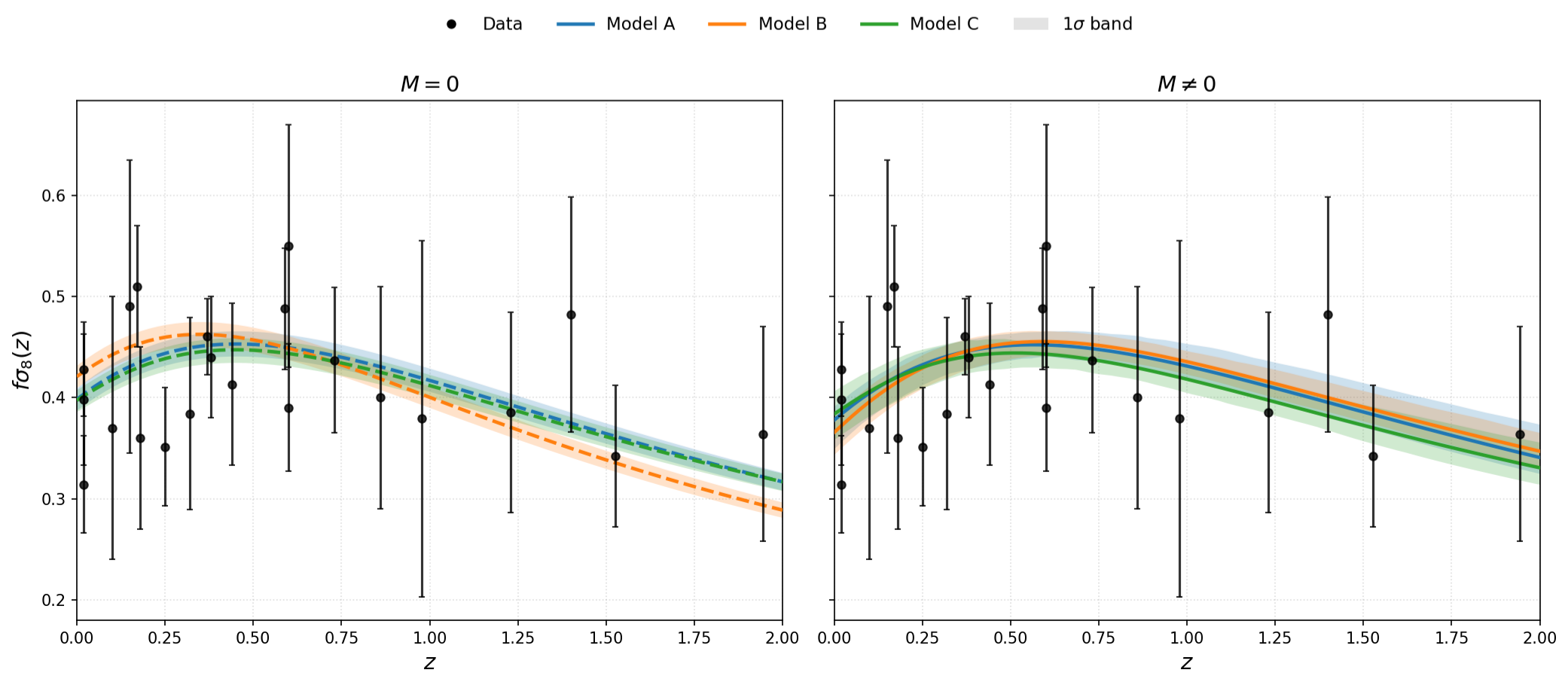}
    \caption{{\it{The evolution of $f\sigma_8(z)$ with $1\sigma$ uncertainty band derived from the posterior distribution.}}}
    \label{fig:fs8}
\end{figure*}

In this section we perform the observational confrontation as described above, and we discuss the obtained results.  For all scenarios considered we summarize the marginalized posterior mean values in Table~\ref{tab:result} and information criteria in Table~\ref{tab:property}.  Furthermore,  in Fig.~\ref{fig:corner} we present the posterior distributions of the free parameters. In addition, in Fig.~\ref{fig:fs8} we show the redshift evolution of $f\sigma_8$ predicted by the marginalized mean parameters of each model, together with the $1\sigma$ uncertainty band derived from the posterior distribution. This provides a direct comparison with the observational data and with the $\Lambda$CDM prediction at the level of observables. Let us now analyze the constraints on $\sigma_8$ and $M$ for each background model separately, and discuss their statistical interpretation and physical significance.

\begin{table}[htp]
\renewcommand{\arraystretch}{1.5}
	\begin{center}
    \makebox[\textwidth][c]{
		\begin{tabular}{ccccccccc}
			\hline
			\hline
			Model & \multicolumn{3}{c}{$M=0$} & \multicolumn{4}{c}{$M$ free} \\
\cmidrule(r){2-4} \cmidrule(l){5-8}
 &  $\sigma_8$ & $\chi_{min}^2$ & $\Delta \chi_{min}^2$ & $\sigma_8$  & $M$ & $\chi_{min}^2$ & $\Delta \chi_{min}^2$ \\ \hline
			A    & $0.784^{+0.022}_{-0.022}$  & $24.33$ & $0$ 
               & $0.842^{+0.043}_{-0.052}$  & $138.55^{+71.577}_{-123.819}$ & 
$22.99$ & $-1.34$  \\
			B    & $0.743^{+0.021}_{-0.021}$  & $31.50$ & $7.17$ 
			   & $0.855^{+0.034}_{-0.034}$  & $322.710^{+90.673}_{-140.154}$ & 
$22.91$ & $-1.42$\\
            C    & $0.788^{+0.022}_{-0.023}$ & $23.39$  & $-0.94$  
			    & $0.822^{+0.040}_{-0.049}$  & $84.678^{+66.086}_{-111.259}$ & 
$23.06$ & $-1.27$ \\
			\hline
			\hline
		\end{tabular}}
	\end{center}
    	\caption[]{Marginalized posterior mean values with 68\% credible intervals for each model. Since $M$ has units of $[L^{-1}]$, i.e. of $H$, for convenience we measure it in $ (km\cdot s^{-1}\cdot Mpc^{-1})$. The definitions of Models A, B, and C follow equation \eqref{eq:model} and all $\Delta \chi_{min}^2$ values are computed relative to $\Lambda$CDM.}
        \label{tab:result}
\end{table}

\begin{table}[htp]
\renewcommand{\arraystretch}{1.5}
	\begin{center}
    \makebox[\textwidth][c]{
		\begin{tabular}{ccccccccc}
			\hline
			\hline
Model & \multicolumn{4}{c}{$M=0$} & \multicolumn{4}{c}{$M$ free} \\
\cmidrule(r){2-5} \cmidrule(l){6-9}
 &  AIC$_c$ & BIC & $\Delta \mathrm{AIC}_c$ & $\Delta \mathrm{BIC}$ &  AIC$_c$ & BIC & $\Delta \mathrm{AIC}_c$ & $\Delta \mathrm{BIC}$ \\
\hline
 A & 26.48 & 27.66 & 0   & 0  & 27.47 & 29.65 & 0.99  & 1.99  \\
 B & 33.65 & 34.83 & 7.17  & 7.17  & 27.39 & 29.58 & 0.91  & 1.92\\
 C & 25.54 & 26.72 & -0.94 & -0.94 & 27.54 & 29.73 & 1.06  & 2.07  \\
			\hline
			\hline
		\end{tabular}}
	\end{center}
    	\caption[]{The information criteria AIC$_c$ and BIC for the examined cosmological models, along with the corresponding differences to $\Lambda$CDM.}
        \label{tab:property}
\end{table}

\subsection{Model A: $\Lambda$CDM background}

Model~A reproduces the standard $\Lambda$CDM background expansion and therefore 
serves as a baseline reference.  
When $M=0$, the RSD data prefer a slightly lower clustering amplitude than 
Planck, with $\sigma_8$ falling $1.18\sigma$ below the Planck 2018 value 
($\sigma_8 = 0.811\pm 0.006$). This represents a statistically mild deviation, consistent 
with the well-documented tendency of low-redshift probes to favor weaker 
structure growth.

Allowing the square-root term to vary shifts the inferred $\sigma_8$ upward, 
bringing it fully into agreement with Planck within $0.65\sigma$. However, the resulting improvement in the fit is not statistically significant, as indicated by the small $\Delta$AIC$_c$ and $\Delta$BIC values, which do not provide meaningful support for the inclusion of $M$. Thus, within a $\Lambda$CDM background, the perturbation-only modification introduced by $M$ is not required by current data, although it may accommodate the mild suppression of growth suggested by RSD measurements.

\subsection{Model B: Hubble-tension alleviating background}

The behavior of Model~B is qualitatively different. When constrained to $M=0$, 
this model predicts a significantly lower $\sigma_8$, lying $3.11\sigma$ below 
the Planck value. This reflects the fact that background modifications designed 
to increase $H_0$, as Model~B does, typically reduce the growth rate, thereby 
intensifying the $\sigma_8$ tension when considered alone.

However, introducing the square-root term changes this picture significantly. 
Once $M$ is allowed to vary, the preferred value of $\sigma_8$ rises and 
becomes entirely consistent with Planck at the $1.27\sigma$ level. In parallel, the fit quality improves substantially. Both AIC$_c$ and BIC decrease by more than seven units ($\Delta \mathrm{IC} > 7$), suggesting that the inclusion of $M$ is favored, although part of the improvement is naturally attributable to the additional freedom introduced by the extra parameter.

This result makes the key conceptual advantage of the $\sqrt{Q}$ 
correction clear: it can restore compatibility with Planck-level clustering 
amplitudes in models whose background evolution would otherwise strongly 
suppress structure growth. In this sense, Model~B with $M\neq 0$ represents a 
highly successful example of using perturbative freedom to rescue an 
otherwise disfavored background scenario. This is one of the main results of 
the present work.

\subsection{Model C: Quintom dark energy background}

Model~C, based on a quadratic $f(Q)$ form capable of generating a quintom-like 
equation of state, performs comparably to, or even slightly better than, the 
$\Lambda$CDM-based Model~A. When $M=0$, the preferred value of $\sigma_8$ 
already lies close to the Planck result within $1.01\sigma$ level.

Allowing $M$ to vary does not alter this conclusion: the posterior distribution peaks near $M=0$, indicating that the data do not favor an additional perturbative degree of freedom. However, the inclusion of $M$ further reduces the deviation of $\sigma_8$  from the Planck measurement to a mere $0.24\sigma$. We emphasize that, although Model~C contains multiple background parameters, these are fixed to the pre-fitted best-fit values from \cite{Yang:2025mws}, such that the background evolution is not treated as flexible in our perturbation analysis. This result therefore suggests that the perturbation structure implied by this specific background evolution is already sufficient to accommodate RSD measurements, without requiring further modifications to $G_{\rm eff}$.

Nonetheless, as shown in~\cite{DESI:2024mwx}, this model does not resolve 
the Hubble tension once background observables are included. In 
particular, DESI BAO combined with 
Pantheon+ and CMB data yield $H_0 = 68.03 \pm 0.72\ {\rm 
km\,s^{-1}\,Mpc^{-1}}$, which remains in tension with local distance ladder 
measurements.

\begin{table}[htbp]
    \centering
    \begin{tabular}{lcccc}
        \hline\hline
        Model
        & \multicolumn{2}{c}{$\Delta\mathrm{ln}~\mathcal{Z}$} 
        \\
        & $M=0$ &  $M$ free  \\
        \hline
        Model A & $0$& $6.267$ \\
        Model B & $ -3.607$ & $3.395$ \\
        Model C & $0.199$ & $5.656$ \\
        \hline\hline
    \end{tabular}
    \caption{Differences in Bayesian evidence relative to $\Lambda$CDM.}
    \label{tab:bayesian evidence}
\end{table}

\subsection{Physical interpretation and parameter degeneracies}

Let us now discuss the aforementioned results. As we can see, the square-root 
correction $M\sqrt{Q}$ is theoretically notable because it modifies the growth 
of structure while leaving the background expansion unaltered. Our results show 
that a positive value of $M$ suppresses the growth rate, equivalently reducing 
$G_{\rm eff}$, and thereby compensates for the low-redshift preference for a 
suppressed $f\sigma_8$. This allows the intrinsic value of $\sigma_8$ to remain 
closer to the Planck-inferred amplitude without compromising the agreement with 
RSD data.

Because the analysis uses RSD alone, the inferred reduction of the $\sigma_8$ tension remains limited by the $M-\sigma_8$ degeneracy inherent in $f\sigma_8$ measurements, as we observe from the posterior contours in Fig.~\ref{fig:corner}, Increasing $\sigma_8$ enhances the 
clustering amplitude, while increasing $M$ suppresses the growth through a 
reduced effective gravitational strength. Their effects therefore counterbalance 
each other, producing elongated likelihood contours. This degeneracy explains 
why $\sigma_8$ can shift towards higher values (appearing more consistent with 
Planck) when $M$ is included: the perturbative modification absorbs the growth 
suppression required by low-redshift observations.

To further quantify the statistical preference for this extended parameter space, we compute the Bayesian evidence for all considered models. The resulting differences in log-evidence relative to the $\Lambda$CDM model are summarized in Table~\ref{tab:bayesian evidence}. Although the Bayesian evidence formally favors the extended model (with $\Delta \ln \mathcal{Z} \sim 6$), this result must be interpreted with caution. As previously noted, the improvement in the maximum goodness-of-fit is marginal. The enhanced evidence is therefore primarily driven by the aforementioned $M-\sigma_8$ degeneracy: along the degenerate direction the likelihood does not decrease, which effectively suppresses the Occam penalty.  Consequently, this enhanced evidence may not constitute strong support for new physics.

We should mention here that breaking this degeneracy will require incorporating 
complementary probes such as weak-lensing shear, CMB lensing, or full-shape 
galaxy spectra, all of which respond to $G_{\rm eff}$ and the growth history in 
different ways. Current RSD data  are insufficient on their 
own to fully disentangle these effects.

In summary, our analysis confirms that the $\sqrt{Q}$ correction provides a 
minimal and physically motivated perturbative mechanism that can help alleviate
the $\sigma_8$ tension across a wide range of background cosmologies, with 
especially strong impact in models that simultaneously aim to address the 
Hubble tension.

 \section{Conclusions}
\label{sec:conc}

The persistent discrepancy between the clustering amplitude inferred from 
early-universe probes such as Planck and that measured by low-redshift 
large-scale structure surveys, continues to motivate the exploration of 
extensions to general relativity. Within this context, we investigated a class 
of symmetric teleparallel gravity models of the form $f(Q)=F(Q)+M\sqrt{Q}$, 
where the square-root correction introduces a new perturbative degree of 
freedom that modifies the effective gravitational coupling $G_{\mathrm{eff}}$ 
while leaving the background expansion history
strictly unchanged in FLRW background. This feature 
makes the model a theoretically appealing and minimal candidate for 
addressing the $S_8$ tension, as it isolates structure growth modifications 
from the background cosmology.

Using the latest set of redshift-space distortion measurements, including DESI 
DR1 Full-Shape constraints, we carried out a systematic parameter estimation 
analysis across three representative background scenarios: (i) a $\Lambda$CDM 
background (Model~A), (ii) a modified expansion designed to alleviate the 
Hubble tension (Model~B), and (iii) a quintom dark-energy background motivated 
by recent DESI reconstructions (Model~C). Our   findings can be summarized 
as follows.

For Model~A ($\Lambda$CDM background), we were expecting that  taking $M=0$  we 
recover completely the standard cosmological paradigm, which already shows only 
a mild ($1.18\sigma$) deviation from the Planck value of $\sigma_8$, consistent 
with known low-redshift trends. Introducing the $M\sqrt{Q}$ correction leads to a marginal improvement in the agreement, bringing $\sigma_8$ within $1\sigma$ of Planck, though this improvement is not statistically significant.
Although the statistical preference for $M\neq 0$ is modest, this scenario 
demonstrates that the perturbative correction can fine-tune growth dynamics 
without altering the background.

In the case of Model~B (Hubble-tension-oriented background), we found that the 
square-root correction displays its most significant impact in this scenario. 
When $M=0$, the model predicts a value of $\sigma_8$ that deviates from Planck 
by $3.11\sigma$, placing it in strong tension with growth-rate data. 
However, once the square-root term is activated, the discrepancy is fully 
removed, and $\sigma_8$ becomes consistent with Planck at the $1\sigma$ level. 
While both AIC and BIC favor the inclusion of $M$, suggesting that perturbative modifications help recover compatibility with data in this background, this improvement must be attributed primarily to the extra degrees of freedom. This result highlights the capacity of $M\sqrt{Q}$ to 
rescue background cosmologies that would otherwise be disfavored by 
structure formation measurements, and is the main result of this work.

In the case of Model~C (quintom dark energy), the posterior peaks near $M=0$, showing that the baseline 
dynamical background already matches the RSD data without requiring additional 
perturbative freedom. Nonetheless, this scenario does not resolve the Hubble 
tension, consistent with recent Gaussian-process reconstructions based on DESI 
DR2, Pantheon+, and compressed CMB data~\cite{Yang:2025mws}.

Our theoretical analysis confirms that a positive value of $M$ suppresses the 
linear growth of structure by reducing the effective gravitational strength. 
This mechanism can bring the lower clustering amplitude preferred by low-redshift data into closer agreement with the higher amplitude favored by early-time physics. However, the persistent $M$--$\sigma_8$ degeneracy indicates that this effect primarily reflects a redistribution of parameter degeneracies rather than a genuinely predictive explanation at this stage.
Moreover, since our analysis does not include a full set of background datasets, the impact on the $H_0$ tension cannot be robustly assessed. Therefore, rather than providing a definitive or joint resolution of the $H_0$ and $S_8$ tensions, the square-root correction should be regarded as a physically interpretable modification that may help alleviate these tensions within a unified $f(Q)$ framework.

At the same time, the posterior contours reveal an extended degeneracy between $M$ and $\sigma_8$, reflecting the competing effects between enhanced intrinsic clustering and suppressed growth through $G_{eff}$. Its presence reveals the possible limitations of current RSD datasets, which still suffer from relatively large uncertainties, partial inconsistencies among surveys, a notable sensitivity to the choice of fiducial cosmology (especially $\Omega_m$) and unavoidable degeneracy between constraints on growth rate and clustering amplitude. These issues currently hinder a precise determination of $M$ using RSD alone.

Recent developments strengthen this perspective. Studies based on the cosmic 
shear three-point correlation function~\cite{DES:2025llp} and EFTofLSS 
analyses~\cite{DAmico:2025zui} suggest that improved non-linear modeling within 
$\Lambda$CDM may also reduce the tension. This indicates that the origin of the 
$S_8$ anomaly may arise from a combination of late-time gravitational physics, 
non-linear structure formation, and observational systematics. Distinguishing 
between these possibilities is a central open challenge.

We close this work by mentioning that to break the degeneracy between $M$ and 
$\sigma_8$ and decisively test the physical relevance of the square-root 
correction, upcoming work must incorporate complementary probes. In particular, 
weak-lensing measurements, full-shape galaxy clustering, higher-order 
statistics, redshift-space multipoles, and CMB lensing, respond differently to 
modifications of $G_{\mathrm{eff}}$ and to the evolution of matter 
perturbations. Future multi-probe analyses, especially those combining Euclid, 
DESI Year~2+, Rubin LSST, and CMB-S4 data, will dramatically enhance sensitivity 
to perturbative gravitational effects. Regarding cosmological consistency, the results of \cite{Atayde:2021pgb} provide a natural reference point, though their constraints are primarily anchored to 
early-Universe CMB data; the mild deviation from our late-time RSD-based results is therefore not unexpected. A joint analysis incorporating galaxy clustering and weak gravitational lensing measurements is left for future work. In conclusion, considered 
together, these observational and theoretical efforts will be crucial for 
determining whether the $M\sqrt{Q}$ correction constitutes a genuine physical 
signature or an artifact resulting from the current stage of cosmological 
modeling.

\section*{Acknowledgments}
We are grateful to Qingqing Wang,  Yating Peng for insightful comments. This 
work was supported in part by the National Key R\&D Program of China 
(2021YFC2203100, 2024YFC2207500), by the National Natural Science Foundation of 
China (12433002, 12261131497, 92476203), by CAS young interdisciplinary 
innovation team (JCTD-2022-20), by 111 Project (B23042),  by Anhui Postdoctoral 
Scientific Research Program Foundation (No. 2025C1184), by CSC Innovation 
Talent Funds, by USTC Fellowship for International Cooperation, and by USTC 
Research Funds of the Double First-Class Initiative. XR is supported by ``Talent Scientific Fund of Lanzhou University'' and the activity ``APCTP-2026-F02''. ENS acknowledges the contribution of the LISA CosWG and the COST Actions  and of COST Actions CA21136 ``Addressing observational tensions in cosmology with systematics and fundamental physics (CosmoVerse)'',  CA21106 ``COSMIC WISPers in the Dark Universe: Theory, astrophysics and experiments (CosmicWISPers)'', and CA23130 ``Bridging high and low energies in search of quantum gravity (BridgeQG)''.

\appendix

\section{The cancellation of the $\sqrt{Q}$ term in Friedmann equation}
\label{appA}

For the square-root contribution $f(Q)=M\sqrt{Q}$, one has
\begin{equation}
    f_Q=\frac{M}{2\sqrt{Q}}, \qquad f_{QQ}=-\frac{M}{4Q^{3/2}}.
\end{equation}
In a spatially flat FLRW universe, where $Q=6H^2$, its contribution to the first Friedmann equation is
\begin{equation}
    6f_QH^2-\frac{1}{2}f
    =6H^2\frac{M}{2\sqrt{Q}}-\frac{M\sqrt{Q}}{2}
    =\frac{MQ}{2\sqrt{Q}}-\frac{M\sqrt{Q}}{2}
    =\frac{M\sqrt{Q}}{2}-\frac{M\sqrt{Q}}{2}=0.
\end{equation}
Likewise, its contribution to the second Friedmann equation is
\begin{equation}
    12H^2f_{QQ}+f_Q
    =12H^2\left(-\frac{M}{4Q^{3/2}}\right)+\frac{M}{2\sqrt{Q}}
    =-\frac{3MH^2}{Q^{3/2}}+\frac{M}{2\sqrt{Q}}
    =-\frac{M}{2\sqrt{Q}}+\frac{M}{2\sqrt{Q}}=0,
\end{equation}
where in the last step we used $Q=6H^2$. Therefore, the squared-root term cancels identically from both background Friedmann equations.

\bibliographystyle{JHEP}
\bibliography{sqrt}

\end{document}